\documentclass[12pt,cite]{article}
\usepackage{epsfig}
\usepackage{amsmath}
\newcommand {\nut} {$\nu_{\tau}$}

\newcommand{\lsim}{\mathrel{\raisebox{-.6ex}{$\stackrel{\textstyle<}{\sim}$}}}

\begin{document}
\begin{flushright}
\end{flushright}
\vspace{1cm}

\begin{center}

\Large{\bf UHE and EHE neutrino induced taus\\ inside the Earth}

\vspace{1.5cm}

\large{S. Bottai\footnote{Corresponding author.\\ 
Address: Dipartimento di Fisica, Universit\`a di Firenze, 
via G.~Sansone~1, I-50019 Sesto F.no (FI), Italy.\\  
Tel.: +39-0554572254. Fax: +39-0554572121.
E--mail: bottai@fi.infn.it}, S. Giurgola\footnote{E--mail: giurgola@fi.infn.it}}

\vspace{0.5cm}

\normalsize{\it University of Florence and INFN Sezione di Firenze,\\
via G.~Sansone~1, I-50019 Sesto F.no (FI), Italy.}

\vspace{2cm}

\begin{abstract}

Tau neutrinos interacting inside the Earth produce $\tau$ leptons
which thereafter can decay inside the atmosphere.  The propagation of
extremely energetic \nut's and $\tau$'s through the Earth is studied 
by means of a detailed Monte Carlo simulation, taking into account
all major mechanisms of \nut~  
interactions and $\tau$ energy loss as well as decay modes. 
The rates of $\tau$'s emerging from the Earth
are determined as a function of $\tau$'s energy for several cosmic neutrino models.

\vspace{0.5cm}



\end{abstract}

\end{center}

\newpage

\section{Introduction}

\begin{figure}
\begin{center}
\epsfig{figure=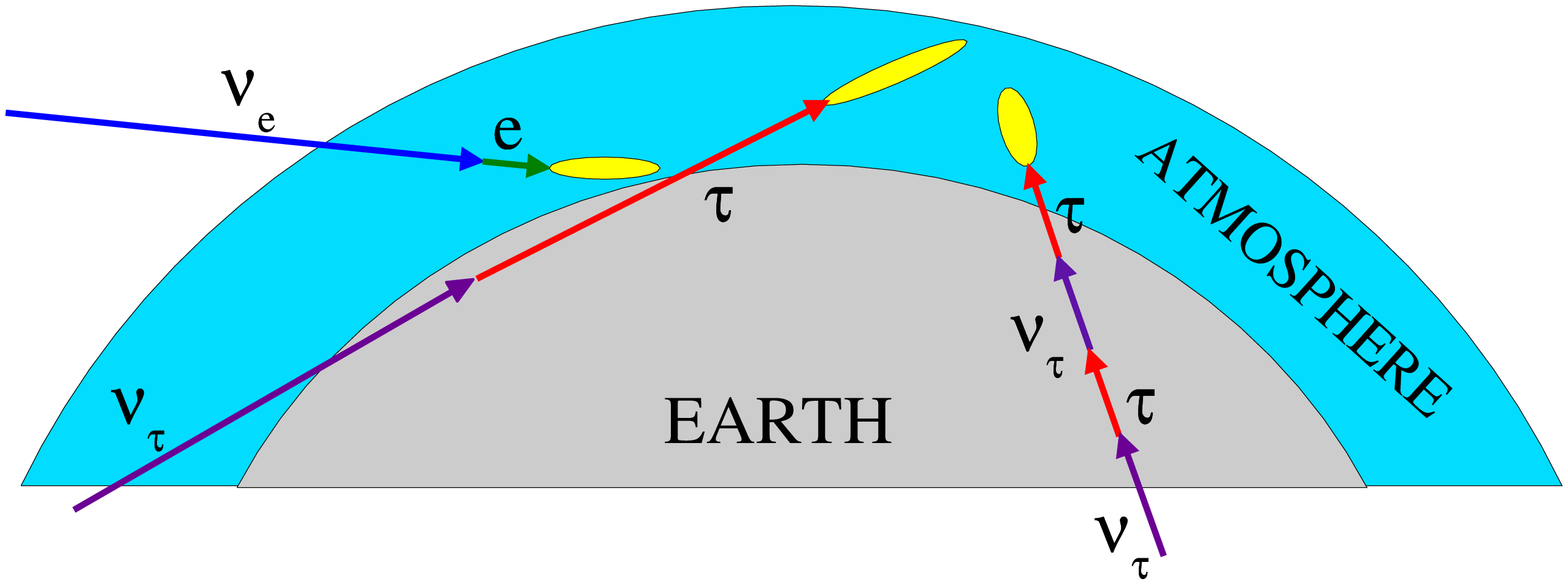, width=15cm}
\caption{Schematic drawing of atmospheric showers induced by neutrinos.
Upward going $\nu_{\tau}$'s could be detected as upward going showers
induced by $\tau$ decay.}
\label{fig:nuart}
\end{center}
\end{figure}

Cosmic neutrinos give a unique opportunity to open
a new observational window in the field of 
Astrophysics and Cosmology. The detection of such particles, not deflected
or absorbed during their intergalactic path, could be revealing for
the identification of sources of Ultra
and Extreme High Energy Cosmic Rays (UHECR and EHECR).
The detection of such ultra and extreme high energy neutrinos is a challenge as it
demands a very large sensitive area. The experiments of the upcoming 
generation plan to use underwater-underice Cherenkov detectors~\cite{gai95} and 
large field-of-view atmospheric fluorescence detectors~\cite{lin97}~\cite{auger}.
One of the best evidence of cosmic neutrinos would be the 
detection of upstream showers/particles emerging from the Earth. For this 
kind of events
the atmospheric muon and primary charged cosmic ray background would be 
completely suppressed. On the other hand this signature can hardly be
observed at extreme energy because the rise of weak cross sections entails 
the opacity of the Earth with respect to neutrino
propagation~\cite{nau98}~\cite{gand}~\cite{bottai}.\\
The problem of muon neutrino propagation 
through the Earth can be solved by taking properly into account neutrino-nucleon
scattering inside the Earth, either by Monte-Carlo simulation 
~\cite{bottai} or by approximate iterative solution of the relevant transport 
equation~\cite{nau98}.\\
It has been recently pointed out in references 
\cite{far97}~\cite{hal98}~\cite{bot99}~\cite{Iyer}~\cite{bottai} that the 
behaviour of $\tau$-neutrinos, 
whose existence should be guaranteed in a neutrino-oscillation scenario, 
should be significantly different from $\nu_{\mu}$ and $\nu_e$.
Whilst muon and electron neutrinos are practically absorbed after
one charged current (CC) interaction, the $\tau$ lepton created by the $\nu_{\tau}$ CC
interaction may decay 
in flight before losing too much energy, thereby generating a new $\nu_{\tau}$ with
comparable energy. Hence ultra high energy $\tau$-neutrinos and $\tau$'s should emerge 
from the Earth instead of being absorbed. 
For a correct evaluation of the energy spectrum of $\nu_{\tau}$'s and $\tau$'s emerging
from the Earth, one has to properly take into account $\nu_{\tau}$ interactions
as well as $\tau$ energy loss and decay. An analytical approach similar
to that used in~\cite{nau98} for $\nu_{\mu}$ has been proposed
by S. Iyer \emph{et al.}~\cite{Iyer} 
neglecting $\tau$ energy loss. This approach holds as long as energy
does not exceed $10^{16}$ eV because above this energy $\tau$ interaction
length becomes comparable with $\tau$ decay length~\cite{bottai}.\\ 
In reference ~\cite{bottai} a detailed Monte Carlo calculation of $\nu_{\tau}-\tau$ 
system propagation through the Earth has been performed for energy up to 
$10^{20}$~eV including the $\tau$ energy loss contribution, with special 
emphasis on the initial $\nu_{\tau}$ spectrum deformation as a function of the zenith 
angle of the emerging particle.
As pointed out in ref. ~\cite{fargio2}~\cite{bertou}~\cite{bottai2} the
tau leptons, created by CC $\nu_{\tau}$ interactions inside the Earth,
because of their relative long decay length at high energy,
could emerge from the Earth surface and eventually
decay inside the atmosphere (fig.~\ref{fig:nuart}). Such kind of events could be detected by 
atmospheric shower detectors as upward going showers.
In reference~\cite{fargio2} an overview of the physics processes
involved in case of $\tau$'s emerging from Earth and mountains is presented,
suggesting the
possibility of the detection of such events. In reference~\cite{bertou} the rate of
expected events in the Auger detector is given on the basis of a Monte Carlo
simulation while in reference~\cite{bottai2}, and more extensively in the present work,
the results of a detailed Monte Carlo simulation are presented in a general framework,
giving the rate of emerging $\tau$'s for unit Earth surface above a given
minimum energy, from $10^{14}$~eV to $10^{22}$~eV and for a large variety of
neutrino fluxes. The results are presented
considering two different calculations of $\tau$ photonuclear cross section and
comparing directly the rate of emerging $\tau$ events to the rate of Extensive
Air Showers (EAS)
produced by neutrinos interacting inside the atmosphere.

\section{Lepton interactions}

\begin{figure}
\begin{center}
\epsfig{figure=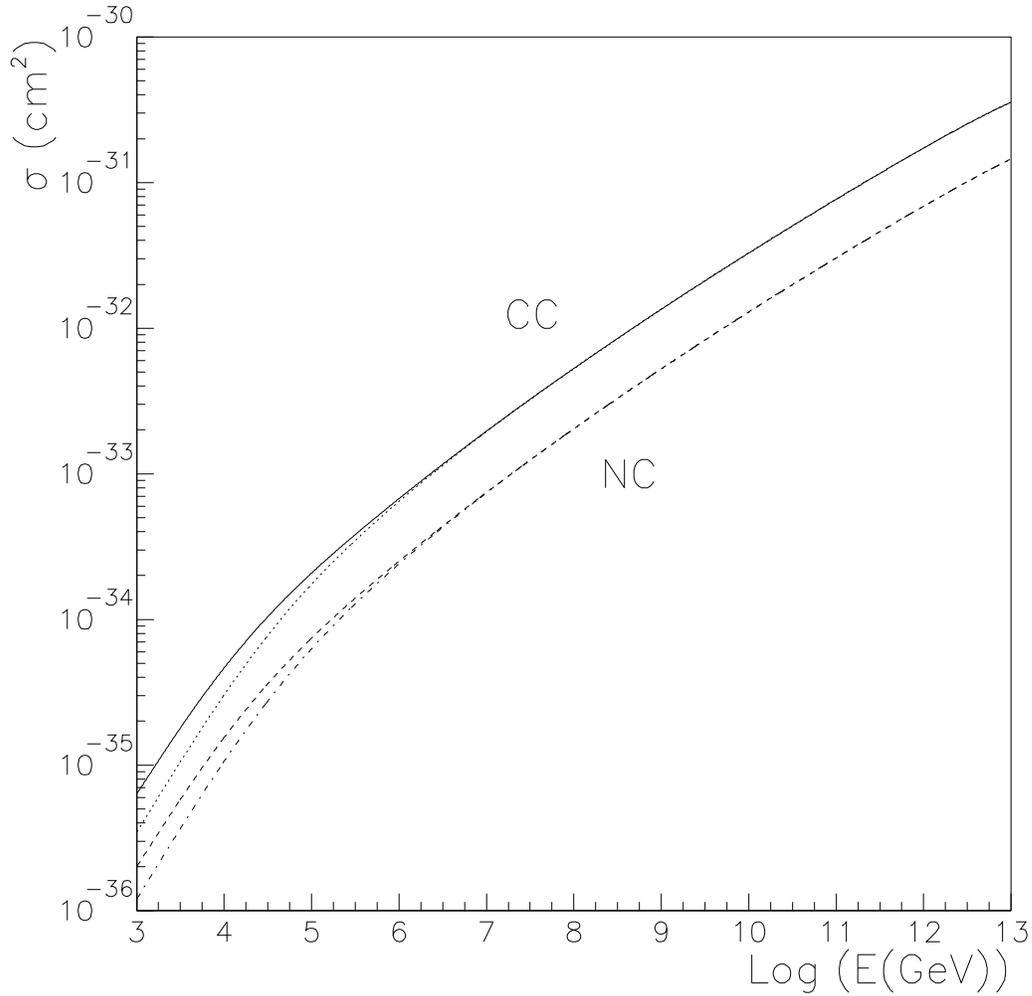, width=15cm}
\caption{Total $\nu_{\tau}$ CC and NC cross sections on isoscalar nucleon target
as a function of energy. Continuos line shows the neutrino CC cross section; dotted line
shows the antineutrino CC cross section; dashed line shows the neutrino NC cross section and
dotted-dashed line shows the antineutrino NC cross section.
CTEQ3-DIS parton distribution has been used for the calculation.}
\label{fig:sigmanu}
\end{center}
\end{figure}

Deep inelastic neutrino-nucleon scattering is the dominant interaction of energetic 
neutrinos into conventional matter. Charged and neutral current differential cross 
sections~\cite{gand} used for this simulation~\cite{bottai} have been calculated in the
framework of QCD improved 
parton model. 
Since a large contribution to cross sections comes from the very low $x$ region, laying 
beyond present accelerator domain, an extrapolation of parton distribution at very low 
$x$ ($x\approx 10^{-8}$) is necessary. We have used the parton distribution set CTEQ3-DIS~\cite{lai}, 
available in the program library PDFLIB~\cite{plo97}, with NLO-DGLAP formalism used for 
the evolution with the invariant momentum transfer between the incident neutrino and 
outgoing lepton ($-Q^2$) and assuming for very low $x$ the same functional form measured at 
$x=\mathcal{O}$($10^{-5}$). Even if more sophisticated approaches for low $x$ extrapolation 
have been developed using dynamical QCD~\cite{glu98} the results for cross section 
calculations (fig.~\ref{fig:sigmanu}) do not differ more than $10$\% from the approach 
taken here with CTEQ3-DIS plus ``brute force'' extrapolation~\cite{gand}~\cite{glu98}.\\
The electromagnetic interactions of muons with matter, at the considered energies, are dominated by radiative processes 
rather than ionisation~\cite{lipa}. Cross sections for electromagnetic
radiative processes of $\tau$ are lower than muon's, yet radiative 
interactions remain the dominant process for $\tau$ energy loss.
The cross sections used for radiative interactions of
$\tau$ leptons are based on QED calculation for
bremsstrahlung~\cite{pet68} and for 
direct pair production~\cite{kok70} while for photonuclear (PH)
interactions references ~\cite{bez81}~\cite{reno} have been used. For all
aforementioned processes 
we have implemented stochastic interactions for $\nu=(E_{i}-E_{f})/E_{i}\geq 10^{-3}$ 
and a continuous energy loss approximation for $\nu=(E_{i}-E_{f})/E_{i}\le 10^{-3}$, where $E_{i}$ 
and $E_{f}$ are $\tau$ energies before and after the interaction respectively.\\
It is worth mentioning that bremsstrahlung cross section scales as the inverse 
square of lepton mass $m_l$ whereas direct pair production and photonuclear
interaction cross sections approximately scale according 
to $1/m_l$~\cite{reno}~\cite{tan91}. As a consequence the dominant processes for
$\tau$ lepton 
energy loss are direct pair production and even more photonuclear interaction
rather than bremsstrahlung photon radiation~\cite{reno}.
The calculation
of cross sections in case of PH interactions
has been explicitely performed by Bezrukov and Bugaev ~\cite{bez81} only for muons 
while the recent calculation performed by Iyer~\emph{et al.}~\cite{reno} 
includes the $\tau$ lepton. The two
calculations, as far as $\tau$ lepton is concerned, differ by almost one order
of magnitude at extreme energy (fig.~\ref{fig:beta}) and therefore the
rate of emerging $\tau$'s is expected to be sizely affected.
The $\tau$ decay has been simulated by using the TAUOLA package~\cite{jad93}.
The Monte Carlo simulation has been performed following neutrinos and
charged leptons along their path inside the Earth. The Earth model
considered is the preliminary Earth model of ref.~\cite{dzi81}, applied to a
perfectly smooth Earth surface with the outer layer made either of
``standard rock'' or water. For what concerns the Earth composition we have used ``standard rock''
($Z=11,\,A=22$) for the crust and the mantle and iron ($Z=26,\,A=55.8$) for the core.
\section{Neutrino induced $\tau$'s inside the Earth}

\subsection{Effective depth}
Tau neutrinos interacting in proximity of Earth's surface can produce $\tau$'s
which thereafter may survive until they get into the atmosphere. 
The production of $\tau$'s which emerge from the Earth is driven by two, not completely
separated, mechanisms. The first is the propagation of neutrinos through the Earth, 
which determines the amount and the energies of neutrinos approaching 
the crust just below the Earth's surface.
The second is the production and propagation of $\tau$'s in the crust just below the Earth's surface.
In the next section we will show the results of the simulation concerning the whole ``story'' of propagation
of leptons, while in this section we focus the attention only on the second mechanism.
Here we want to evaluate the thickness of the Earth's Crust to be considered active for
production of emerging $\tau$'s as a function of the energy of the $\tau$ neutrino as it
approaches the emerging surface. Hence no propagation through the whole Earth is considered in
this section.
The results given here can be useful also for calculations of the rate of horizontal $\tau$'s
emerging from mountains.

As a first
approximation, the $\tau$ decay length $\gamma \beta c\tau$ can be used to estimate the 
{\it effective depth}, defined as the thickness of Earth's Crust to be considered
as a source for such emerging $\tau$'s. 

\begin{figure}
\begin{center}
\epsfig{figure=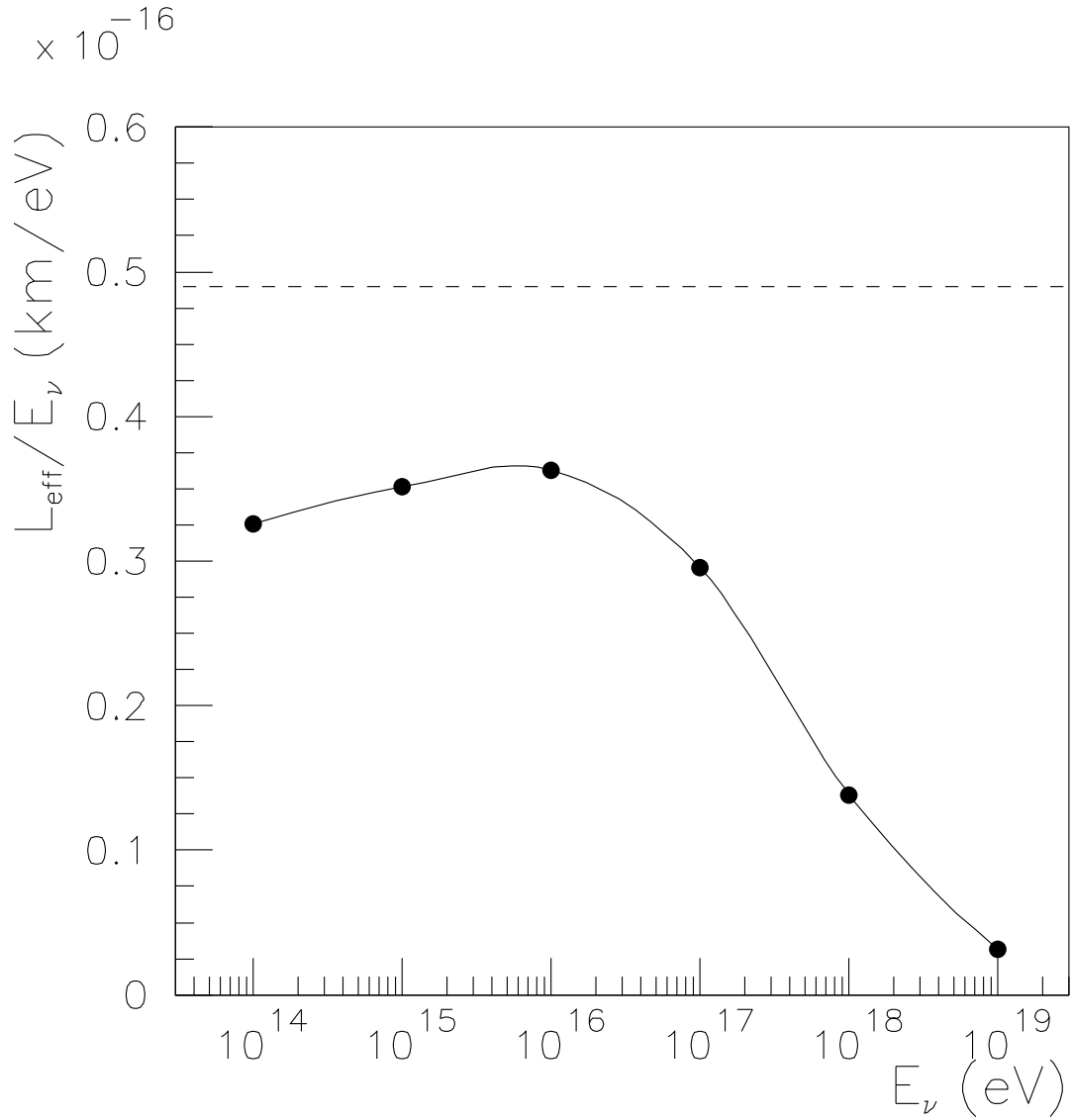, width=15cm}
\caption{{\it Effective depth} for generation of emerging $\tau$'s (PH cross section from
ref.~\cite{bez81} and outer Earth layer made of standard rock). The dotted line
shows the expected result for no interacting $\tau$'s created with the same neutrino energy
and decaying after one $\tau$ decay length.}
\label{fig:leffsue}
\end{center}
\end{figure}

An accurate calculation of the {\it effective depth} has to properly take into account the differential
cross sections describing the neutrino
interactions and tau energy loss and decay. The total neutrino cross section influences the number of
produced $\tau$'s but not the {\it effective depth}.
We have simulated a number $N_{int}^{\nu}$ of $\nu_{\tau}$ CC interactions induced by vertical neutrinos
and uniformly distributed along a column of Earth's Crust within a depth $L$. 
One can then define the {\it effective depth} as the limit:
$$
 L_{eff} =\lim_{L\rightarrow \infty }  \, L \, \epsilon (L) 
$$
where $\epsilon (L)$ is the number of emerging $\tau 's$ divided by the number of simulated
$\nu$ interactions $N_{int}^{\nu}$. 
The {\it effective depth}
is shown in fig.~\ref{fig:leffsue} where it is compared with the
$\tau$ decay length with $E_{\tau}=E_{\nu}$. 
At low energy, $L_{eff}/E_{\nu}$ slowly increases with $E_{\nu}$ because the fraction of
energy carried
by the $\tau$ increases with energy in the CC $\nu_{\tau}$ interactions.
Above $\approx 10^{17}$eV the effect of $\tau$ energy loss becomes relevant and the energy degradation
of $\tau$'s prevents a sizeable fraction of them from reaching the surface.
Consequently, $L_{eff}/E_{\nu}$ drops and $L_{eff}$ roughly approaches
the value of
$\tau$ radiation length in the outer layer of Earth's Crust ($\approx 10$~km for the highest energies and 
PH in ref.~\cite{bez81}).

\subsection{Effective aperture}

In this section the calculation of the fluxes of $\tau$'s emerging
from the Earth has been performed by means of a full simulation of 
$\nu_{\tau}-\tau$ system propagation through the Earth.

\begin{figure}
\begin{center}
\epsfig{figure=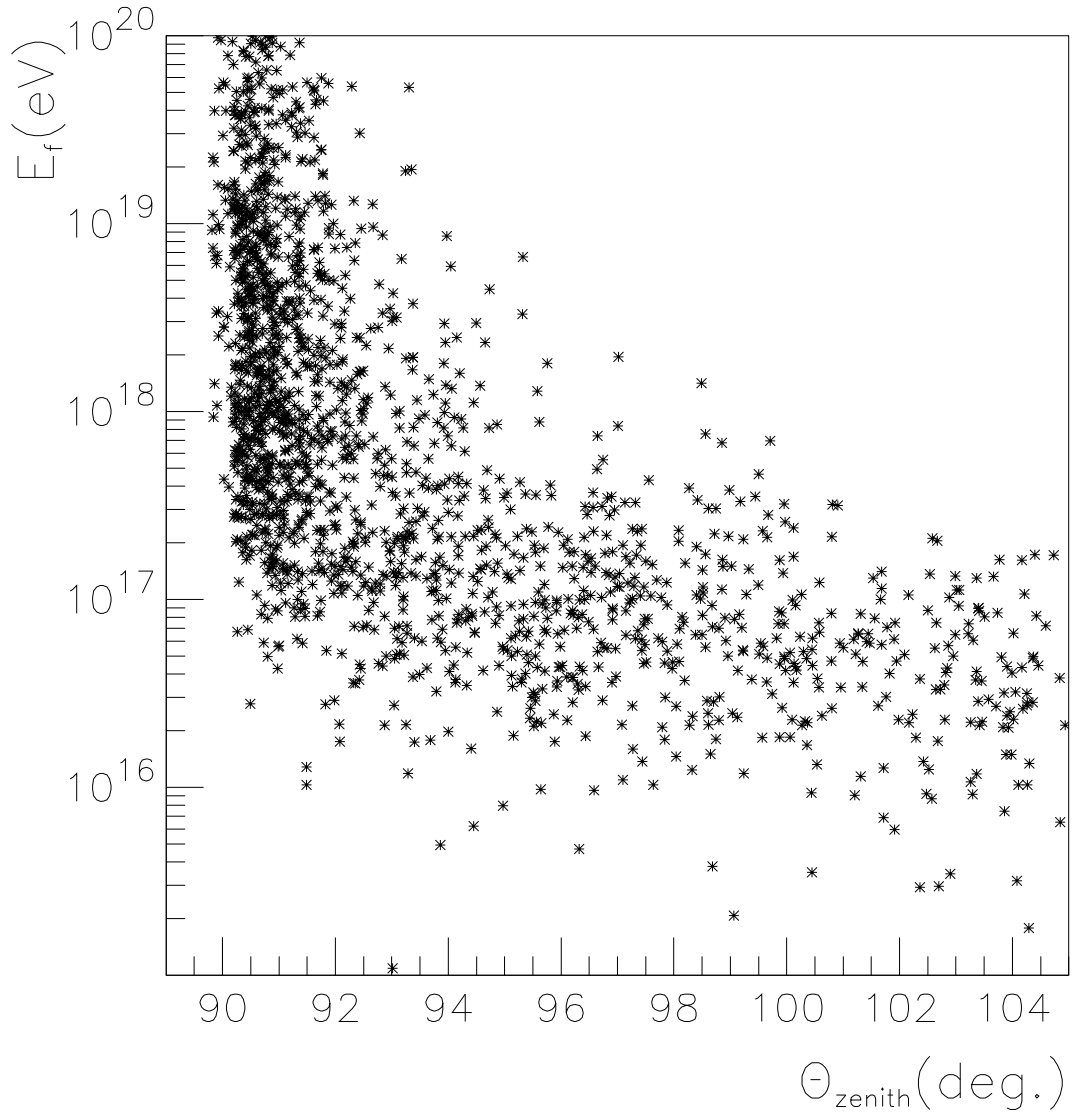, width=15cm}
\caption{Scatter plot of final energies of emerging $\tau$'s ($E_{f}$) versus emerging zenith angles.
The simulation has been performed at $E_{\nu}=10^{20}$eV for an isotropic flux. PH cross section is from
ref~\cite{bez81} and the standard rock is considered for the outer Earth layer.}
\label{fig:nu20}
\end{center}
\end{figure}

Fig.~\ref{fig:nu20} shows
the result of the simulation for an isotropic flux of $\nu_{\tau}$'s
impinging on the Earth with an energy $E_{\nu}=10^{20}$eV. 
Most $\tau$'s with zenith
angles larger than $94^0$ emerge with energies $10^{16}$eV-$10^{17}$eV for which, assuming
PH interaction in ref.~\cite{bez81}, $\tau$
decay length
becames comparable with its radiation length.

For zenith angles corresponding to a path
length in the Earth larger than $\nu_{\tau}$ interaction length in the Crust
($L^\nu _{int}\approx 70$~km for $E_{\nu}=10^{20}$eV), neutrinos
interact far away
from the Earth surface, and the produced
$\tau$'s lose energy and decay before
getting out. Owing to this mechanism, the energy of the $\nu_{\tau}$-$\tau$ system decreases
as zenith angle increases (fig.~\ref{fig:nu20}) and the probability for a $\nu_\tau$ to emerge as a 
$\tau$ drops at large zenith angles.
For $\nu_{\tau}$
energy lower than $E_\nu = 10^{20}$eV the behaviour of the $\nu_{\tau}$-$\tau$ system is similar
but shifted
to larger values of zenith angles. 

The number
of emerging $\tau$'s with energy larger than $E_{min}$ (minimum energy) from an Earth infinitesimal surface $dS$
can be calculated in a unidimensional approach as :
\begin{equation}
dN_{\tau}(E_{\tau}\geq E_{min}) =  \int_{E_{\nu}\geq E_{min}} {\frac{d^2N}{dE_{\nu}\,d\Omega}(E_{\nu})
 \, P_{\nu_{\tau}\rightarrow \tau}(cos(\Theta),E_{\nu},E_{min})\, dS\, |cos(\Theta)|\,
 d\Omega\, dE_{\nu}}
\label{eqn:flu1}
\end{equation} 
where $\frac{d^2N}{dE_{\nu}\,d\Omega}(E_{\nu})$ is the cosmic $\tau$ neutrino flux,
$\Theta$ is the zenith angle of the emerging particles, $P_{\nu_{}\rightarrow \tau}
({\rm cos}(\Theta),E_{\nu},E_{min})$ is the probability that a $\nu_{\tau}$ impinging on the
Earth surface will emerge as a $\tau$ with an energy $E_{\tau}\geq
E_{min}$. Assuming a spherical shaped Earth and an isotropic neutrino flux, Eq.~\ref{eqn:flu1} 
turns into :
\begin{equation}
N_{\tau}(E_{\tau}\geq E_{min}) =  S\, \int_{E_{\nu}\geq E_{min}} {\frac{d^2N}{dE_{\nu}\,d\Omega}(E_{\nu})}
 \, \int P_{\nu_{\tau}\rightarrow \tau}(cos(\Theta),E_{\nu},E_{min})\,
 |cos(\Theta)|\, d\Omega \, \,
 dE_{\nu}
\label{eqn:flu2}
\end{equation} 
where $S$ is a portion of Earth surface.
 
One can now define an
{\it effective aperture} $A_{eff}(sr)$ containing
all the physics of leptons propagation :
$$
A_{eff}(E_{\nu},{E_{min}})= \int{P_{\nu_{\tau}\rightarrow \tau}(cos(\Theta),E_{\nu},E_{min})
\, |cos(\Theta)|\, d\Omega  
}
$$
so that  Eq.~\ref{eqn:flu2} reads :
\begin{equation}
N_{\tau}(E_{\tau}\geq E_{min}) =  S\, \int_{E_{\nu}\geq E_{min}} {\frac{d^2N}{dE_{\nu}\,d\Omega}(E_{\nu})}
 \, A_{eff}(E_{\nu},{E_{min}})\, 
 dE_{\nu}
\label{eqn:convo}
\end{equation} 
The {\it effective aperture} $A_{eff}$ does not depend on neutrino fluxes or detector performance
and allows a simple and straightforward calculation of the number of emerging
$\tau$'s for any cosmic neutrino flux and detector sensitive area.
It must be pointed out that in the definition of $P_{\nu_{\tau}\rightarrow \tau}
({\rm cos}(\Theta),E_{\nu},E_{min})$ we also constrain the emerging $\tau$ to decay within
an atmospheric slant depth larger than 1000~g/cm$^2$. This condition guarantees the
full size formation of an EAS after $\tau$ decay.

\begin{figure}
\begin{center}
\epsfig{figure=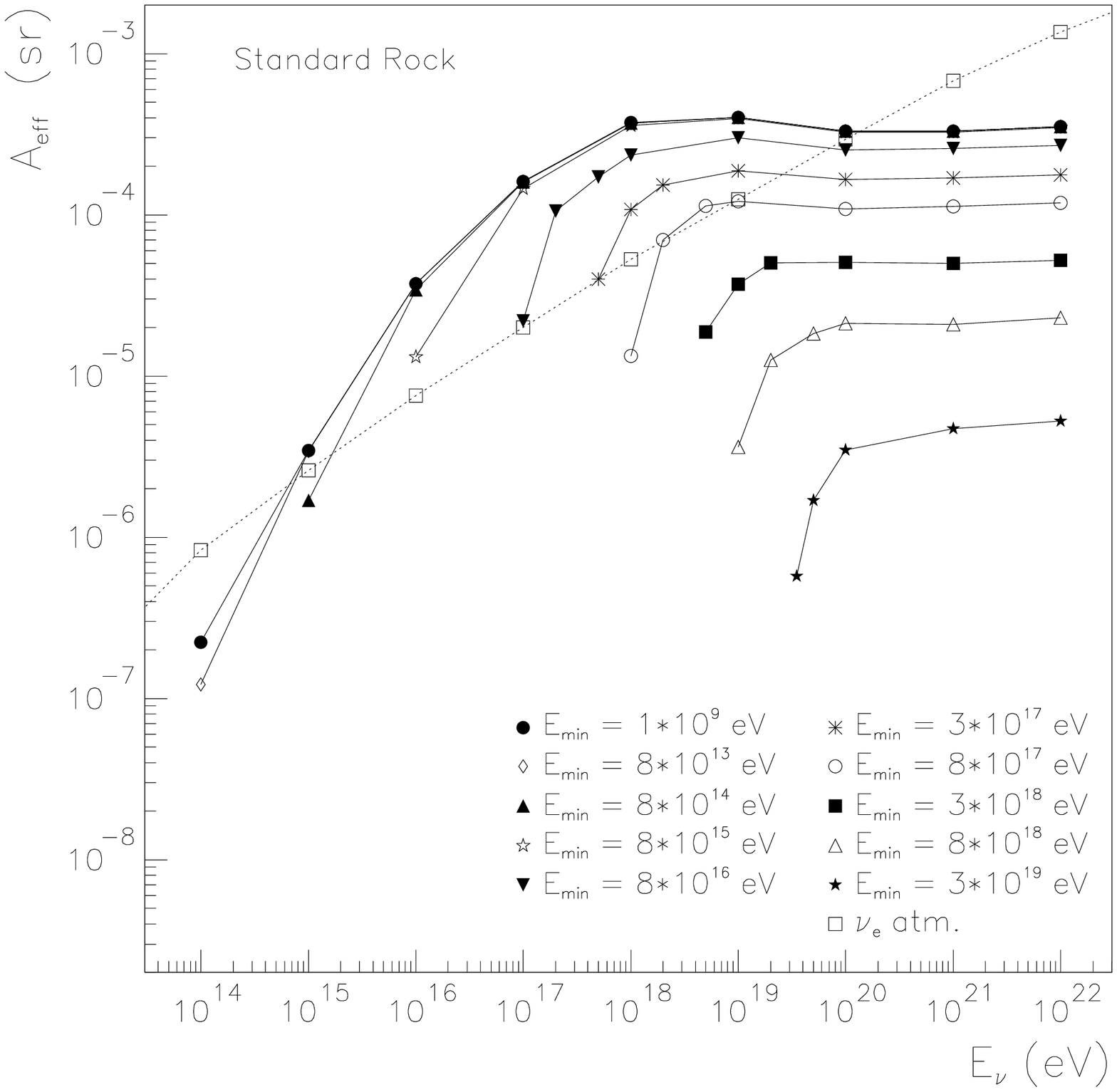, width=15cm}
\caption{{\it Effective aperture} for different
minimum energy $E_{min}$ and standard rock
considered for the outer Earth layer. PH cross section is from ref~\cite{bez81}.
Dotted line shows the {\it effective aperture} for events induced by downward going
$\nu_e$ CC interactions in the atmosphere.}
\label{fig:aeff}
\end{center}
\end{figure}

\begin{figure}
\begin{center}
\epsfig{figure=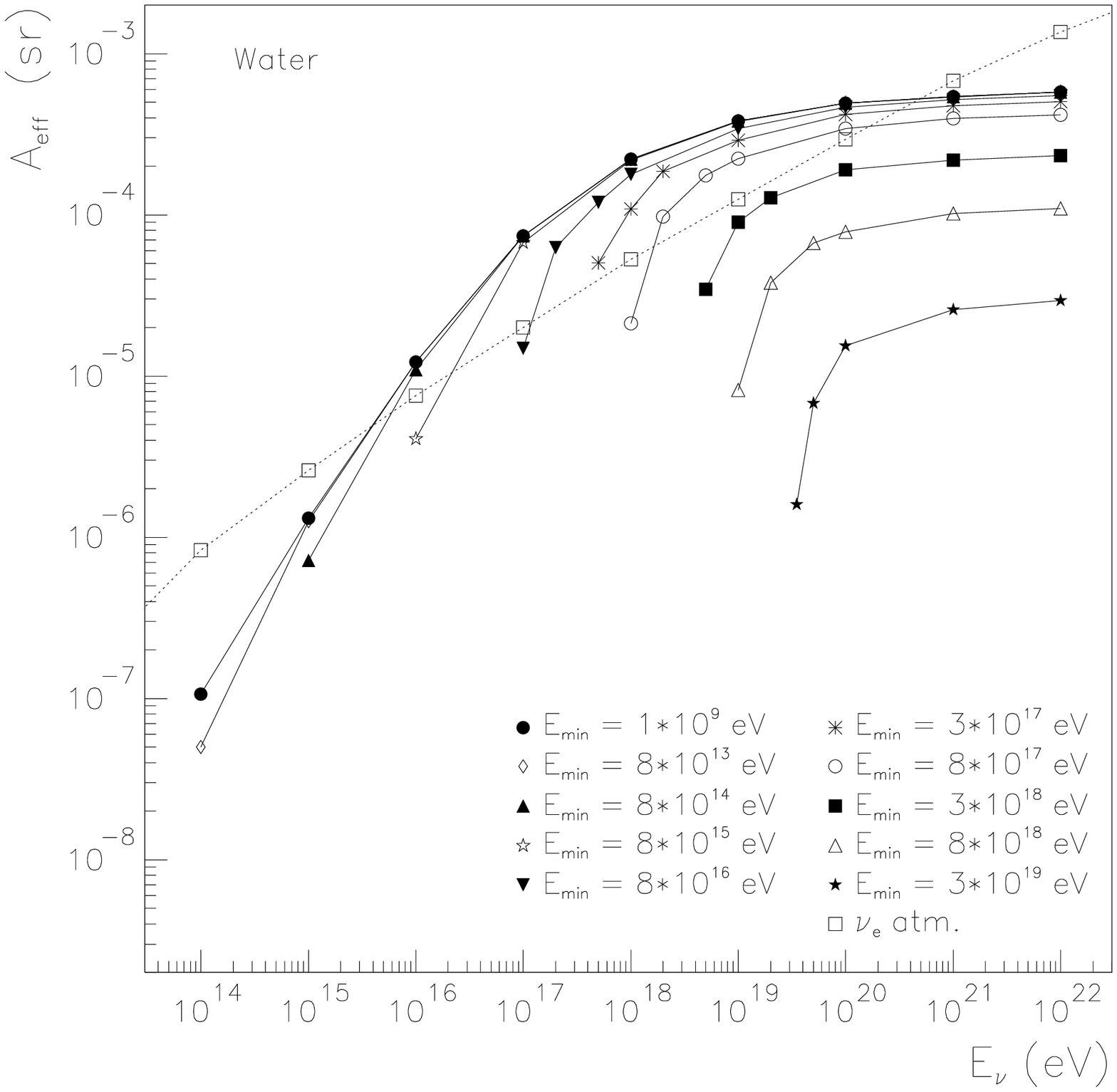, width=15cm}
\caption{{\it Effective aperture} for different
minimum energy $E_{min}$ and 3 km water
considered for the outer Earth layer. PH cross section is from ref~\cite{bez81}.
Dotted line shows the {\it effective aperture} for events induced by downward going
$\nu_e$ CC interactions in the atmosphere.}
\label{fig:aeffsea}
\end{center}
\end{figure}

The results for the {\it effective aperture}
by using Bezrukov and Bugaev PH cross section
are shown in figs.~\ref{fig:aeff} and ~\ref{fig:aeffsea}.
They seem to be in fairly good agreement with those in ref.~\cite{bertou} for the
Auger detector, though the contribution of the effective Auger sensitivity
can not be completely unfolded from quoted results using available informations.

An {\it effective aperture} $A^{\nu_e}_{eff}$ for downwards going $\nu_e$ CC interactions
in the atmosphere is also shown in figs.~\ref{fig:aeff} and ~\ref{fig:aeffsea}.
For this kind
of events, the energy of produced EAS is the same as $\nu_e$. For an isotropic neutrino flux,
the number of EAS induced by $\nu_e$ CC interactions within the atmospheric mass overhanging a
portion of Earth
surface $S$ is given by :
$$
N^{\nu_e}_{EAS}(E_{EAS}\geq E_{min}) =   S\, \int_{E_{\nu}\geq E_{min}} {\frac{d^2N}{dE_{\nu}\,d\Omega}(E_{\nu})
 \, \sigma_{CC}(E_{\nu}) \, N_A \, p \, 2 \pi \, 
 dE_{\nu}}
$$
where $p$ is the atmospheric pressure (column density) at ground level, $N_A$ is the Avogadro number,
$\frac{d^2N}{dE_{\nu}\,d\Omega}(E_{\nu})$ is the cosmic electron neutrino flux and
$\sigma_{CC}(E_{\nu})$ is the neutrino CC cross section on isoscalar target. 
Again, we can define an {\it effective aperture} :

$$
A^{\nu_e}_{eff}(E_{\nu})=2 \pi\, N_A \, p \, \sigma_{CC}(E_{\nu}) 
$$
 
such that the number of $\nu_e$ CC interactions can be written as:
$$
N^{\nu_e}_{EAS}(E_{EAS}\geq E_{min}) =  S \, \int_{E_{\nu}\geq E_{min}} {\frac{d^2N}{dE_{\nu}\,d\Omega}(E_{\nu})
 \, A^{\nu_e}_{eff}(E_{\nu})\, dE_{\nu}}
$$

$A^{\nu_e}_{eff}(E_{\nu})$ increases
with energy following the rise of neutrino cross section. At low energies ($E_{\nu} \ll 10^{18}$eV)
the rise with energy of $A_{eff}$
for emerging $\tau$'s is steeper than the rise of $A^{\nu_e}_{eff}(E_{\nu})$ because both the increase
of $\nu$ cross section and the increase of
$\tau$ decay length contribute to enlarge $P_{\nu_{}\rightarrow \tau}
(cos(\Theta),E_{\nu},E_{min})$.  
$A_{eff}(E_{\nu})$ keeps increasing as long as the increase of $\nu$ cross section
enhances the production of $\tau$'s in the Earth.
Around $E_{\nu}\simeq 10^{18}$eV
the cross section of
neutrinos is so high that the probability for neutrinos to have at least one interaction inside
the Earth is close to one for most zenith angles and the point of first $\nu_{\tau}$ interaction
in the Earth is likely to be far from
the emerging point (this latter effect is responsible for the slow decrease of $A_{eff}$ between
$10^{19}$eV and $10^{20}$eV in fig~\ref{fig:aeff}). 
Moreover, the increase of $\tau$ decay length with energy prevents some of
the extreme high energy events from producing a detectable shower into the atmosphere
(only $\tau$'s decaying within $1000$~g/cm$^2$ slant depth are retained). 
Due to these mechanisms, the value of $A_{eff}$ at extreme high energies
shows an almost flat dependence on energy and strongly depends on
the required $E_{min}$. 

\begin{figure}
\begin{center}
\epsfig{figure=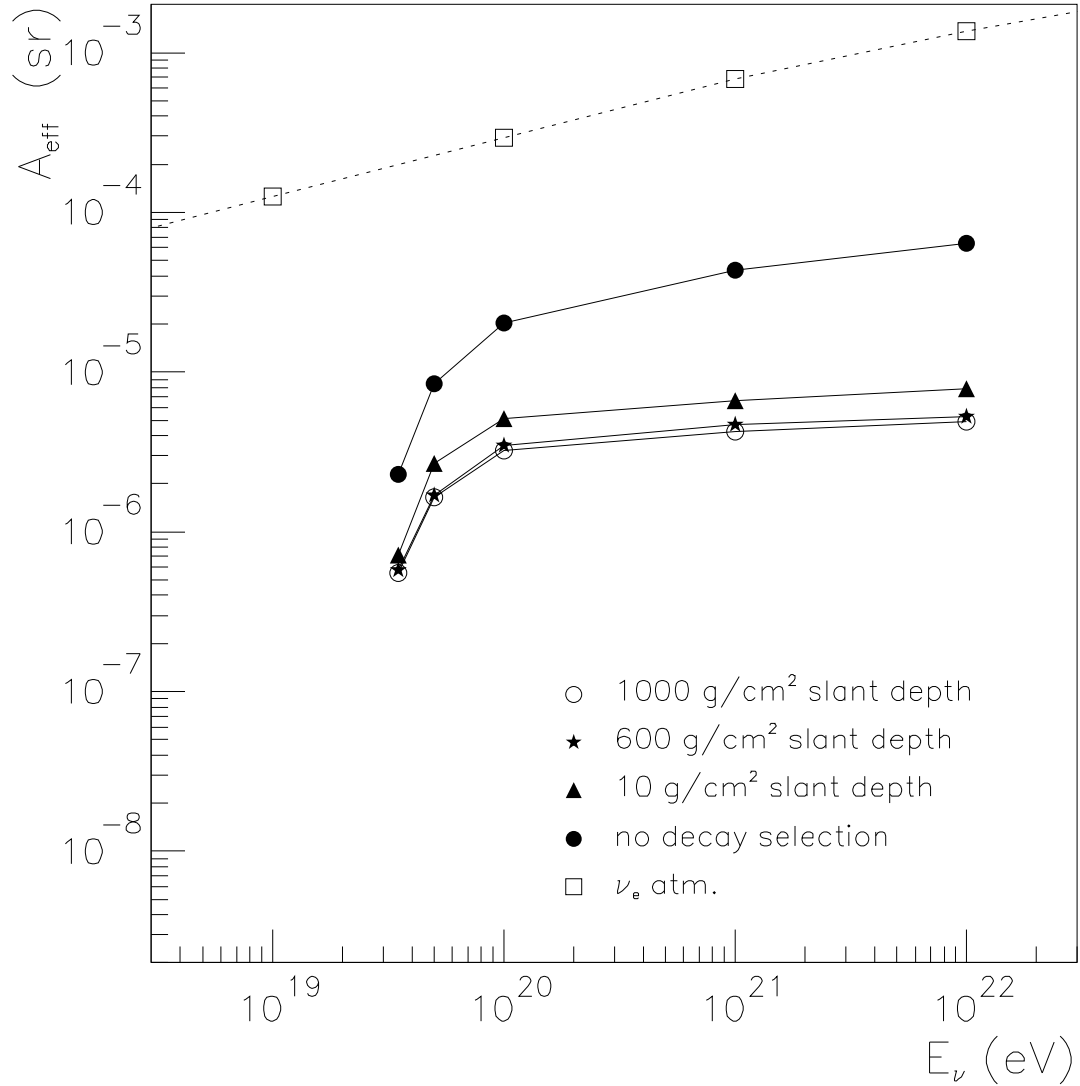, width=15cm}
\caption{{\it Effective aperture} for
minimum energy $E_{min}=3 \times 10^{19}$eV, standard rock
considered for the outer Earth layer and PH cross section from ref~\cite{bez81}.
The effect of different selections on the slant depth of the decay point is shown.}
\label{fig:taudec}
\end{center}
\end{figure}

The effect of decay length selection
at extreme energies is shown in fig~\ref{fig:taudec} while
fig.~\ref{fig:aeffsea} shows the $A_{eff}$ in case of an outer layer of Earth's Crust
made of water and $3$ km thick. Due to the reduction
of target mass, the values of $A_{eff}$ in fig.~\ref{fig:aeffsea} are lower, at low energies, than those
in fig.~\ref{fig:aeff}.
For energies $E_{\nu}\geq 10^{18}$eV, where $A_{eff}$ mainly depends on $\tau$ radiation
length in the outer Earth's Crust, the values in fig.~\ref{fig:aeffsea} are larger 
than those in fig.\ref{fig:aeff}.

\begin{figure}
\begin{center}
\epsfig{figure=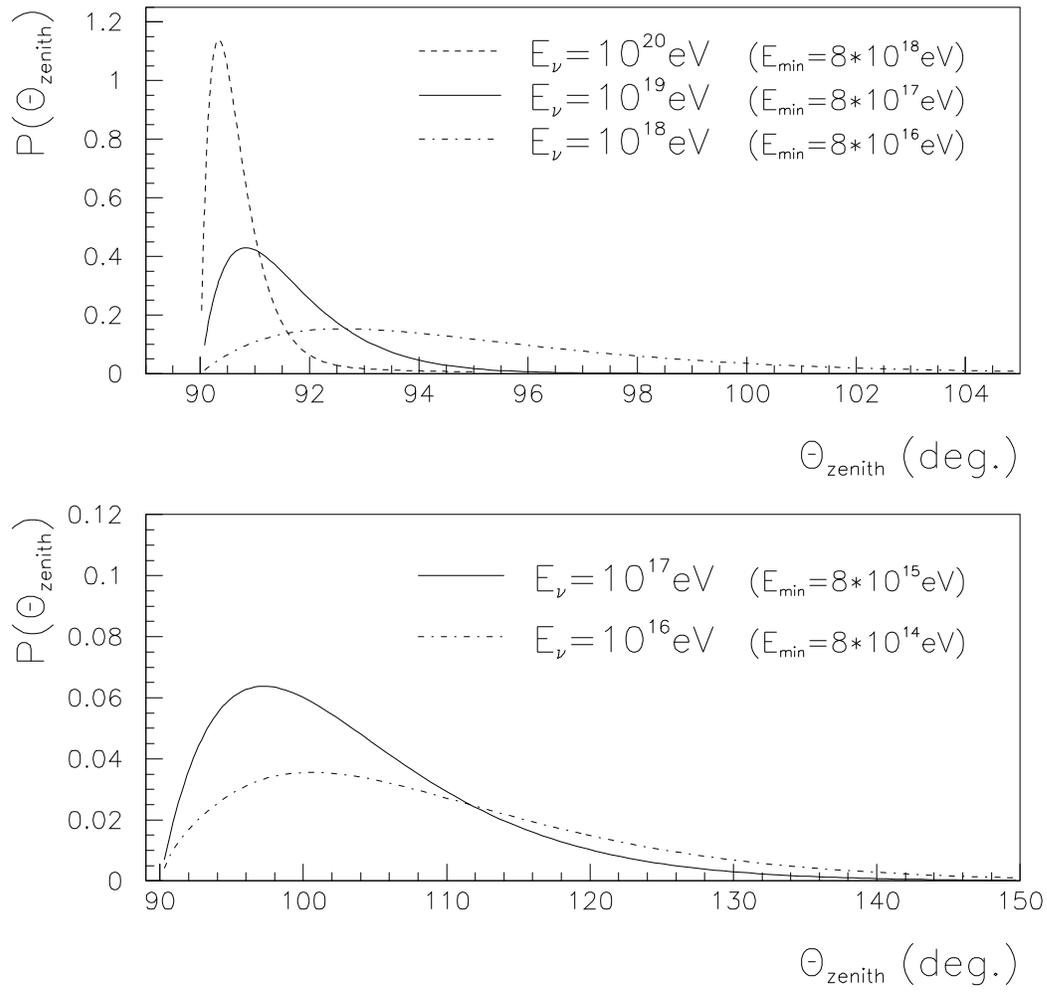, width=15cm}
\caption{Angular distribution of emerging $\tau$'s for several primary neutrino
energies and several minimum energy $E_{min}$.
PH cross section is from ref~\cite{bez81} and standard rock is considered
for the outer Earth layer.}
\label{fig:dis}
\end{center}
\end{figure}

The angular distributions of emerging $\tau$'s are shown in fig.~\ref{fig:dis}.
As the initial neutrino energy increases, the maximum of the angular distribution moves
towards $90^0$ zenith angle.

\begin{figure}
\begin{center}
\epsfig{figure=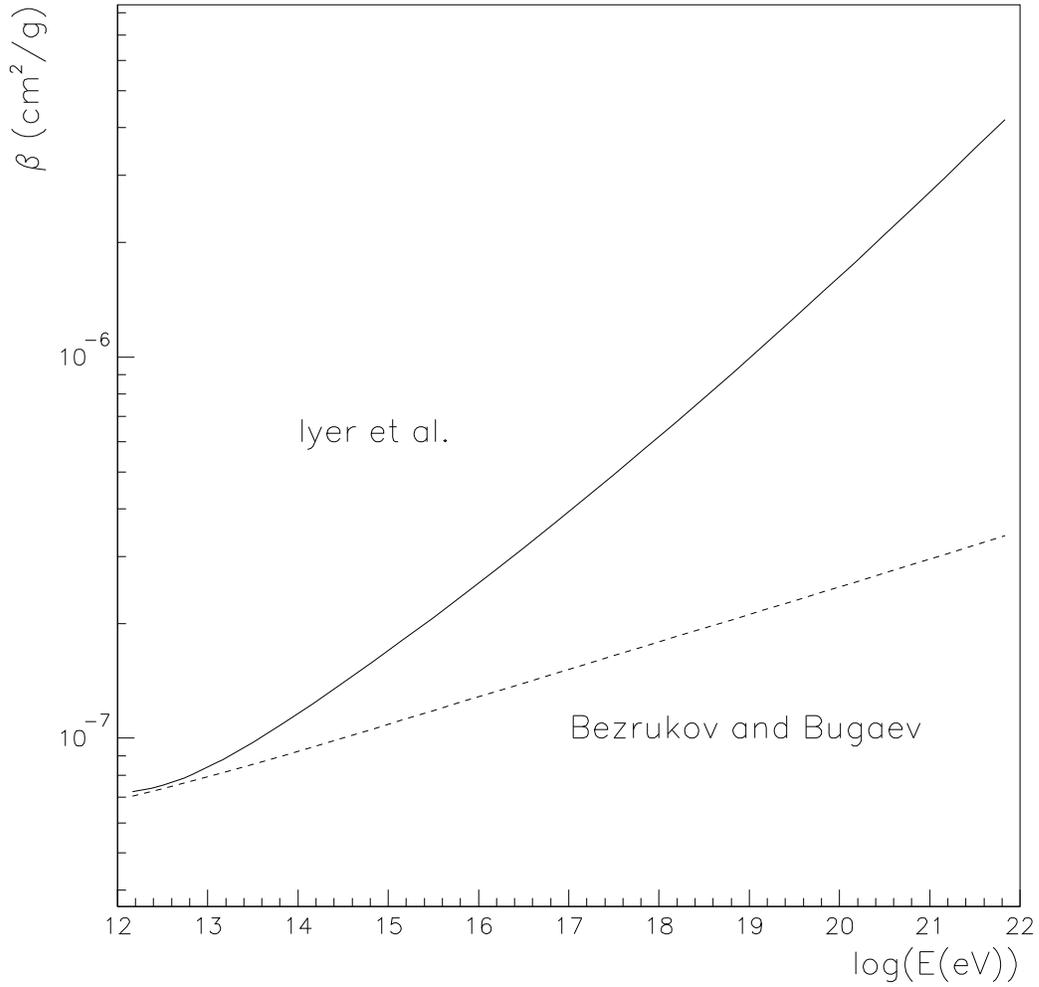, width=15cm}
\caption{$\beta=-\frac{dE/dX}{E} \, \frac{1}{\rho}$ in case of standard rock (A=22, Z=11)
for $\tau$ PH interactions by S. Iyer~\emph{et al.}~\cite{reno} and as it comes out using
formulae of double differential cross section of Bezrukov and Bugaev~\cite{bez81}
with $\tau$ mass.}
\label{fig:beta}
\end{center}
\end{figure}

\begin{figure}
\begin{center}
\epsfig{figure=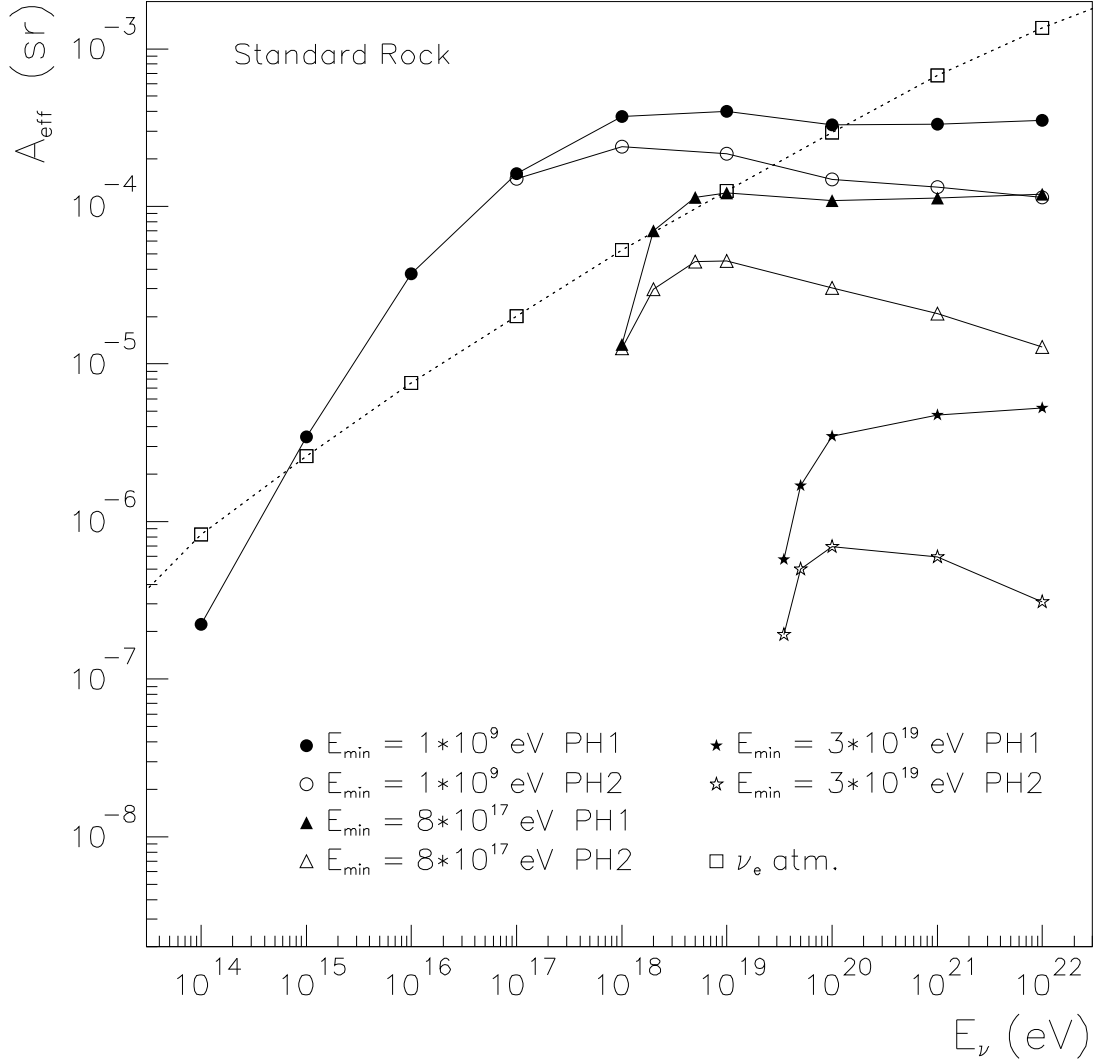, width=15cm}
\caption{{\it Effective aperture} for different
minimum energy $E_{min}$ and standard rock considered
for the outer Earth layer. Calculations using Bezrukov and Bugaev $\tau$
PH cross section~\cite{bez81} (curves labeled as PH1) are
compared with calculations using S. Iyer at al. cross section~\cite{reno}
(curves labeled as PH2).
Dotted line shows the {\it effective aperture} for events induced by downward going
$\nu_e$ CC interactions in the atmosphere.}
\label{fig:aeff2}
\end{center}
\end{figure}

In fig.~\ref{fig:aeff2}, the {\it effective apertures} calculated using
the recent evaluation of $\tau$ PH cross section given in ref.~\cite{reno} are compared
with the {\it effective aperture} of fig.~\ref{fig:aeff}. The higher value of
$\tau$ energy loss (fig~\ref{fig:beta})
gives rise to a suppression of emerging $\tau$'s for the highest energies. 
Moreover, $A_{eff}$ decreases at extreme neutrino energy because 
of the increase with energy of $\tau$ PH radiation length according to fig.~\ref{fig:beta}.

\subsection{Fluxes of emerging $\tau$'s}
According to the hypothesis
of $\nu_\mu\,-\nu_\tau$ neutrino oscillations suggested by the Super\-Kamiokande
experiment~\cite{kamio}, half of the original cosmic $\nu_{\mu}$'s appear as
$\nu_{\tau}$'s after their intergalactic path towards the Earth~\cite{Iyer2}.

Within this scenario, we have used the same fluxes for parent $\nu_{\tau}$ and $\nu_e$: 
$\Phi_{\nu_{\tau}}=\Phi_{\nu_{e}}=\Phi^{initial}_{\nu_{\mu}}/2$ and
we have calculated the rate of emerging $\tau$'s by using 
equation~(\ref{eqn:convo}).

\begin{figure}
\begin{center}
\epsfig{figure=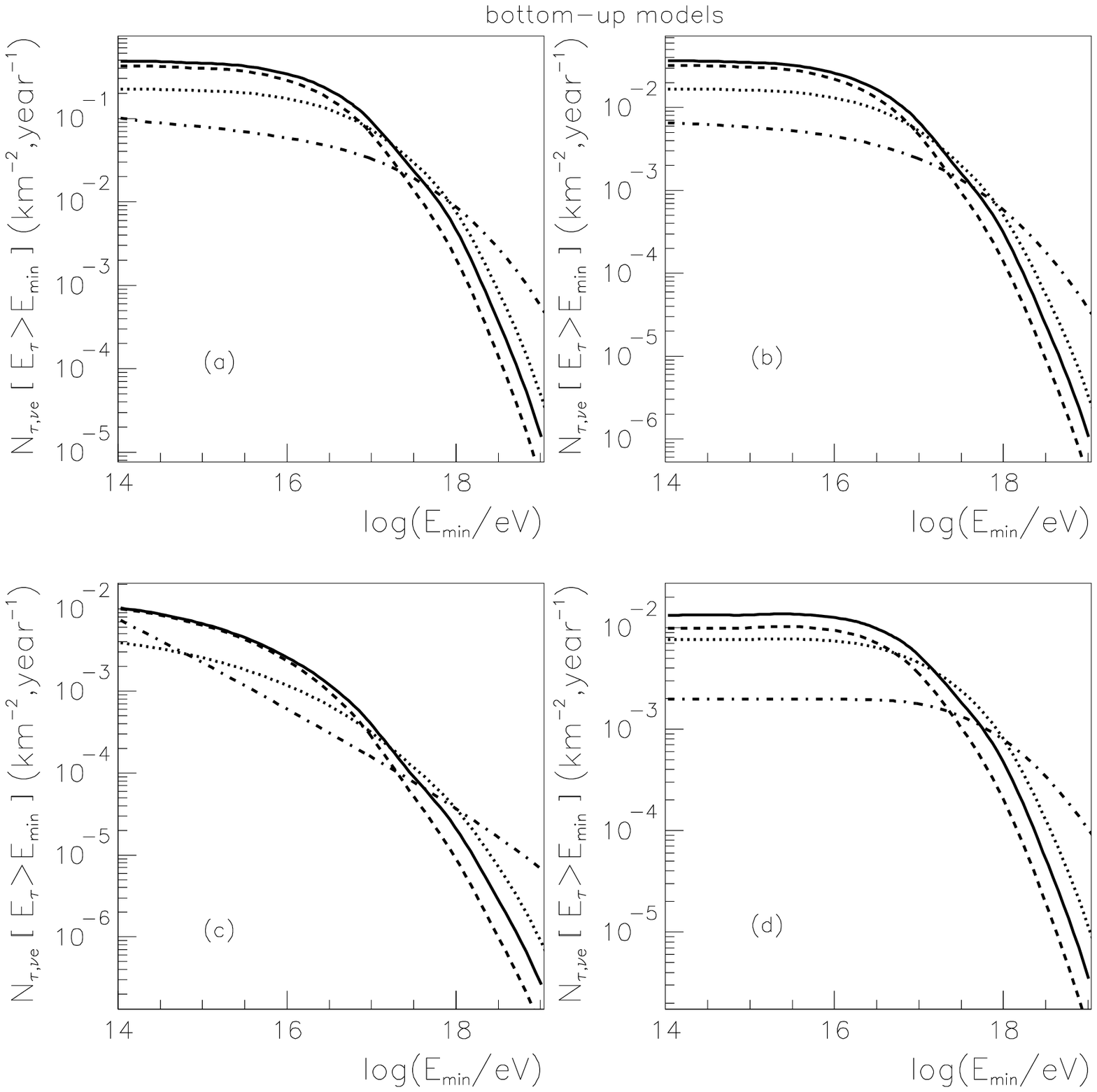, width=17cm}
\caption{Rates of emerging  $\tau$'s versus minimum energy : 
Bezrukov-Bugaev PH cross section~\cite{bez81} and standard rock considered
for the outer Earth layer (continuos lines), Bezrukov-Bugaev PH cross section~\cite{bez81} and
3 km water for the outer Earth layer (dotted lines), S. Iyer \emph{et al.} PH cross section~\cite{reno} 
and standard rock 
for the outer Earth layer (dashed lines). Dashed-dotted lines show the rate of $\nu_e$ CC interactions in the
atmosphere above the unit Earth surface.
Each plot refers to a different \emph{Bottom-Up} model of cosmic neutrino flux:
(a)~AGN (A) from Mannheim~\cite{mannagn}; (b)~AGN (B) from Mannheim~\cite{mannagn};
(c)~GRB from Waxman e Bahcall~\cite{wbgrb} and Vietri extension~\cite{vietri}; 
(d)~GZK neutrinos from Protheroe and Johnson~\cite{prothgzk}.}
\label{fig:bu}
\end{center}
\end{figure}

\begin{figure}
\begin{center}
\epsfig{figure=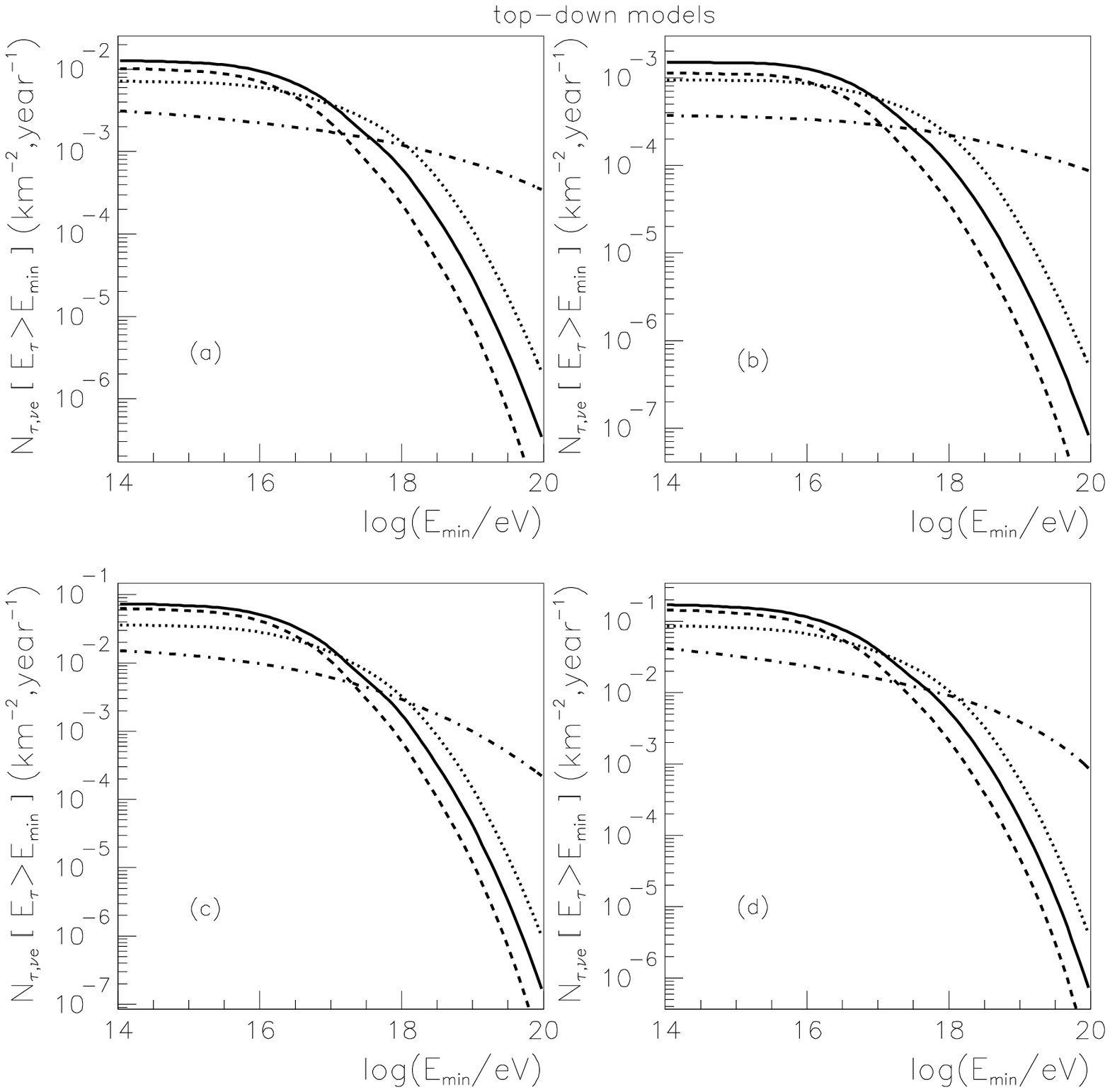, width=17cm}
\caption{Rates of emerging  $\tau$'s versus minimum energy : 
Bezrukov-Bugaev PH cross section~\cite{bez81} and standard rock considered
for the outer Earth layer (continuos lines), Bezrukov-Bugaev PH cross section~\cite{bez81} and
3 km water for the outer Earth layer (dotted lines), S. Iyer \emph{et al.} PH cross section~\cite{reno} 
and standard rock
for the outer Earth layer (dashed lines). Dashed-dotted lines show the rate of $\nu_e$ CC interactions in the
atmosphere above the unit Earth surface. 
Each plot refers to a different \emph{Top-Down} model of cosmic neutrino flux:
(a)~Topological Defect from Protheroe and Stanev~\cite{prothtd}; 
(b)~Topological Defect from Sigl \emph{et al.}~\cite{coppi};
(c)~Topological Defect from Sigl~\cite{sigl2000}; 
(d)~Supermassive Relic Particle from Kalashev \emph{et al.}~\cite{kalashev}.}
\label{fig:td}
\end{center}
\end{figure}

Fig.~\ref{fig:bu},~\ref{fig:td} show the flux of emerging $\tau$'s
compared with the rate of $\nu_e$ CC interactions
in the atmosphere overhanging
the unitary surface.
The signal induced by the decay of emerging $\tau$'s dominate for $E_{min}\lsim 5 \times 10^{17}$eV while
for higher minimum energy $E_{min}$, Earth
matter strongly suppresses the upgoing $\tau$ fluxes. 
It must be emphasized that, in this analysis, we have not taken into account
the difference between induced EAS generated by different $\tau$ decay channels.
It is worth pointing out that leptonic decays
$\tau^- \rightarrow \mu^- \,\bar{\nu_{\mu}} \, \nu_{\tau}$ do not produce any EAS. 

\section{Conclusions}

A detailed study of $\nu_{\tau}$-$\tau$ leptons propagation through the Earth
indicates the likely existence of a quite intense flux of upgoing $\tau$'s
emerging from its surface. The rate of EAS induced by $\tau$'s decay is larger
than the rate of $\nu_e$'s induced EAS for minimum energy $E_{min}\leq 5 \times 10^{17}$eV,
while it drops for larger values of minimum energy.
The rate of emerging $\tau$'s at extreme energies strongly depends on
photonuclear contribution to energy loss, whose calculation is still controversial as it requires 
non-trivial extrapolations. The detection
of such $\tau$  induced events is challenging at a detector's energy threshold $E_{thr}
\approx 10^{19}$eV even for new generation experiments, while for $E_{thr} \ll 10^{18}$eV
the detection seems to be realistic for detectors capable of recording
signals induced by very inclined showers with sensitive areas $\gg 10^3$~km$^2$.


\newpage

\newpage

\newpage

\newpage

\newpage

\newpage

\newpage

\newpage

\newpage

\newpage

\newpage

\newpage

\newpage

\begin{thebibliography}{99}

\bibitem{gai95} T.K. Gaisser, F. Halzen, T. Stanev, Phys. Rep. \textbf{258} (1995) 173 and 
references therein.

\bibitem{lin97} J. Linsley, L. Scarsi, P. Spillantini, Y. Takahashi, Proceedings
XXV ICRC, vol. 5 (1997) 389; ESA and EUSO Team, 
\emph{Extreme Universe Space Observatory - Report on the accommodation of EUSO on
the Columbus Exposed Payload Facility}
ESA/MSM-GU/2000.462/AP/RDA December 2000.
 
\bibitem{auger} Auger Collaboration, \emph{The Pierre Auger Project Design Report}, FERMILAB-PUB-96-024, Jan 1996. 252pp.
 
\bibitem{nau98} V. Naumov, L. Perrone, Astropart. Phys. \textbf{10} (1999) 239 and 
references therein.

\bibitem{gand} Gandhi \emph{et al.},  Astropart. Phys. \textbf{5}  (1996) 81.

\bibitem{bottai} F. Becattini, S. Bottai, Astropart.Phys.  \textbf{15} (2001) 323.  

\bibitem{far97} D. Fargion, \emph{The Role of Tau Neutrino Ultrahigh-energy astrophysics in km$^3$
detectors}, astro-ph/9704205 (1997).

\bibitem{hal98} F. Halzen, D. Saltzberg, Phys. Rev. Lett. \textbf{81} (1998) 4308.

\bibitem{bot99} F. Becattini, S. Bottai, Proceedings XXVI ICRC Salt Lake City 1999, vol. 2 (1999) 249.

\bibitem{Iyer} S.Iyer Dutta, M. Reno, I. Sarcevic, Phys. Rev. D \textbf{62} (2000) 123001

\bibitem{fargio2} D. Fargion, A. Aiello, R. Conversano, Proceedings XXVI ICRC
 Salt Lake City 1999, vol. 2 (1999) 396; D. Fargion astro-ph/0002453, (2000).

\bibitem{bertou} X. Bertou \emph{et al.}, \emph{Tau neutrinos in the AUGER observatory: 
a new window to UHECR sources}, astro-ph/0104452.

\bibitem{bottai2} S. Bottai, S. Giurgola, Proceedings XXVII ICRC Hamburg 2001,
he247, (2001) 1201. 

\bibitem{lai}  H. Lai \emph{et al.}, Phys. Rev. D \textbf{51} (1995) 4763. 

\bibitem{plo97} H. Plothow-Besch, \emph{PDFLIB, Version 7.09}, CERN-PPE 
report 1997.07.02.

\bibitem{glu98} M. Gluck, S.Kretzer and E. Reya, Astropart. Phys. \textbf{11} (1999) 327.


\bibitem{lipa} P. Lipari, T. Stanev, Phys. Rev. D \textbf{44} (1991) 3543.

\bibitem{pet68} A.A. Petrukhin, V.V. Shestakov, Can. J. Nucl. Phys  \textbf{46} (1968) S377.

\bibitem{kok70} R.P. Kokoulin and A.A. Petrukhin, Acta Phys. Hung. \textbf{29} Suppl. 4 (1970) 277.

\bibitem{bez81} L.B. Bezrukov and E.V. Bugaev Sov. J. Nucl. Phys. \textbf{33} (1981) 635.

\bibitem{reno} S.Iyer, M.H. Reno, I. Sarcevic, D. Seckel, Phys. Rev. D \textbf{63} (2001). 

\bibitem{tan91} M.J. Tannenbaum, Nucl. Inst. Meth. A \textbf{300} (1991) 595.

\bibitem{jad93} S. Jadach, Z. Was, R. Decker, J. H. Kuhn, CERN-TH 6793/93 (1993).

\bibitem{dzi81} A.M. Dziewonski and D.L. Anderson, \emph{Physics Earth and Planetary Interiors},
25 (1981) 297.

\bibitem{kamio} S. Fukuda et al., Phys. Rev. Lett. \textbf{85} (2000) 3999. 

\bibitem{Iyer2} S. Iyer, M. H. Reno, I. Sarcevic,  Phys. Rev. D \textbf{62}, 123001 (2000).


\bibitem{mannagn} K. Mannheim, Astropart. Phys. \textbf{3}, 295 (1995).

\bibitem{wbgrb} E. Waxman, J. Bahcall, Phys. Rev. Lett. \textbf{78}, 2292 (1997). 

\bibitem{vietri} M. Vietri, Phys. Rev. Lett. \textbf{80}, 3690 (1998). 

\bibitem{prothgzk} R. J. Protheroe, P. A. Johnson, Astropart. Phys. \textbf{4}, (1995) 253; 
 erratum \textbf{5}, 215 (1996).

\bibitem{prothtd} R. J. Protheroe, T. Stanev, Phys. Rev. Lett. \textbf{77}, (1996) 3708.

\bibitem{coppi} G. Sigl, S. Lee, D. Schramm, P. Coppi, Phys. Lett. B \textbf{392},(1997) 129.
 
\bibitem{sigl2000} G. Sigl,  Lect. Notes Phys. \textbf{556}, (2000) 259.

\bibitem{kalashev} O. E. Kalashev, V. A. Kuzmin, D. V. Semikoz,
 e-print astro-ph/9911035, (1999).
 
 

\end{thebibliography}
\end{document}